\documentclass[%
 reprint,
 amsmath,amssymb,
 aps,
]{revtex4-1}

\usepackage{graphicx}
\usepackage{dcolumn}
\usepackage{bm}
\usepackage{color}
\usepackage{cases}
\usepackage{here}

\begin{document}
\preprint{APS/123-QED}

\title{Phase stability of Au-Li binary systems studied using neural network potential}

\author{Koji Shimizu$^1$}
 \email{shimizu@cello.t.u-tokyo.ac.jp}
\author{Elvis F. Arguelles$^1$}
\author{Wenwen Li$^2$}
\author{Yasunobu Ando$^2$}
\author{Emi Minamitani$^3$}
\author{Satoshi Watanabe$^1$}
  \email{watanabe@cello.t.u-tokyo.ac.jp}

\affiliation{
$^1$Department of Materials Engineering, The University of Tokyo, 7-3-1 Hongo, Bunkyo-ku, Tokyo 113-8656, Japan \\
$^2$Research Center for Computational Design of Advanced Functional Materials, National Institute of Advanced Industrial Science and Technology (AIST), 1-1-1 Umezono, Tsukuba, Ibaraki 305-8568, Japan \\
$^3$Institute for Molecular Science, 38 Nishigo-Naka, Myodaiji, Okazaki, 444-8585, Japan
}

\date{July, 11, 2020}

\begin{abstract}
The miscibility of Au and Li exhibits a potential application as an adhesion layer and electrode material in secondary batteries.
Here, to explore alloying properties, we constructed a neural network potential (NNP) of Au-Li binary systems based on density functional theory (DFT) calculations.
To accelerate construction of NNPs, we proposed an efficient and inexpensive method of structural dataset generation.
The predictions by the constructed NNP on lattice parameters and phonon properties agree well with those obtained by DFT calculations.
We also investigated the mixing energy of Au$_{1-x}$Li$_x$ with fine composition grids, showing excellent agreement with DFT verifications.
We found the existence of various compositions with structures on and slightly above the convex hull, which can explain the lack of consensus on the Au-Li stable phases in previous studies. 
Moreover, we newly found Au$_{0.469}$Li$_{0.531}$ as a stable phase, which has never been reported elsewhere.
Finally, we examined the alloying process starting from the phase separated structure to the complete mixing phase.
We found that when multiple adjacent Au atoms dissolved into Li, the alloying of the entire Au/Li interface started from the dissolved region. 
This paper demonstrates the applicability of NNPs toward miscible phases and provides the understanding of the alloying mechanism.

\begin{description}
\item[DOI]

\end{description}
\end{abstract}

\pacs{Valid PACS appear here}
\maketitle

\section{Introduction}

The Au-Li binary system is known to exhibit various alloy phases over a wide range of compositional ratios \cite{Pelton-BAPD-1986, Bach-EA-2015, Bach-CM-2016, Yang-JACS-2016}.
This remarkable alloy miscibility has recently been attracting attention due to its potential applications, most specifically in next-generation energy storage and electronic devices, among others.
In Li-ion batteries (LIBs), Au is shown to form high-Li-concentration stable alloy phases at low voltages in addition to the low alloying/dealloying potential.
This implies that Au is a promising candidate for an anode electrode material in LIBs \cite{Zeng-FD-2014}.
Moreover, thin-layer Au inserted at the interface of the Li-anode and a solid electrolyte can suppress void formations, resulting in enhanced cyclability and reduction of interface resistance \cite{Kato-JPS-2016, Kato-SSI-2018}.
In novel non-volatile memory devices, the amount of Li at the interface of the Au cathode and Li$_3$PO$_4$ solid electrolyte controls switching behavior \cite{Sugiyama-aplm-2017, Shimizu-PRM-2020}.
These aspects highlight the significance of atomic scale-based studies of Au-Li structural properties and to what extent the alloying proceeds in materials design and development of electrochemical devices.

Experiments using thermal analysis and X-ray measurement report the observed Au-Li phases to be $\alpha$ (17 to 40 at.\% Li), Au$_5$Li$_4$, $\beta$ (47 to 56 at.\% Li), $\delta$ (62.5 to 65.5 at.\% Li), AuLi$_3$, and Au$_4$Li$_{15}$ \cite{Pelton-BAPD-1986}.
Electrochemical measurements and theoretical calculations identify the AuLi$_3$ and Au$_2$Li phases as the stable alloy phases, and Au$_3$Li$_5$, Au$_2$Li$_3$, and Au$_5$Li$_3$ as the metastable phases \cite{Bach-CM-2016}.
Another theoretical work based on density functional theory (DFT) calculations presents Au$_3$Li, AuLi, AuLi$_2$, and AuLi$_3$ as stable phases at 0 GPa \cite{Yang-JACS-2016}.
The available stable phases stored in the Materials Project (MP) \cite{MP} database are Au$_3$Li, AuLi, AuLi$_3$, and Au$_4$Li$_{15}$.
Discrepancies among the previous studies seen above arise from the intrinsic complexity of miscible alloy materials.
Furthermore, high computational costs loaded in a theoretical investigation over a wide compositional space contribute to the lack of consensus on the stable Au-Li phases.

Conventional theoretical studies on alloying properties of other materials also employ DFT calculations to obtain the mixing energies of various compositions and to draw phase diagrams \cite{Luytena-C-2009, Angqvist-PRM-2019, Aspera-JEM-2017}.
However, in most cases, research studies focus on a few stable phases owing to the aforementioned formidable computational costs of a comprehensive search over a wide range of compositions at the $ab~initio$ level.
Alternatively, classical molecular dynamics (MD) simulations using empirical interatomic potentials are able to take into consideration various computational conditions; however, the accuracy of this approach is limited by the quality of the empirical interatomic potentials, which is often insufficient.
In recent years, machine learning (ML) interatomic potentials, e.g., neural network potential (NNP) \cite{Behler-PRL-2007}, Gaussian approximation potential (GAP) \cite{Bartok-PRL-2010}, and spectral neighbor analysis potential (SNAP) \cite{Thompson-JCP-2015}, have been gaining much attention because of their lower computational costs by several order of magnitude while achieving comparable accuracy with DFT calculations.

Particularly, ML potentials have been successfully applied to binary and ternary alloy systems \cite{Hajinazar-PRB-2017, Kobayashi-PRM-2017, Onat-PRB-2018, Li-PRB-2018, Wen-PRB-2019}.
The NNPs of the Cu-Pd, Cu-Ag, Pd-Ag, and Cu-Pd-Ag systems reproduce defect and formation energies and phonon properties well in comparison to DFT calculations \cite{Hajinazar-PRB-2017}, wherein an efficient way of constructing multi-element NNPs are proposed, so-called stratified NN.
The NNPs also have been applied to Al-Mg-Si alloys \cite{Kobayashi-PRM-2017}, Li-Si alloys \cite{Onat-PRB-2018}, and Pd-Si alloys \cite{Wen-PRB-2019}, where all NNPs accurately predict physical properties comparable to DFT results.
Moreover, the Ni-Mo interatomic potential constructed using SNAP has demonstrated that the predicted phase diagram agrees well with the experiment, as well as the DFT-level prediction of various physical properties \cite{Li-PRB-2018}.
The ML potentials have also been applied to simulate Li-ion diffusion in amorphous-Li$_3$PO$_4$ \cite{Li-JPC-2017} and in Li$_{10}$GeP$_2$S$_{12}$ and Li$_7$La$_3$Zr$_2$O$_{12}$ \cite{Marcolongo-CSC-2019}, and to simulate the Li intercalations in carbon \cite{Fujikake-JCP-2018} anodes.
These potentials were able to accurately reproduce their corresponding DFT quantities and exhibited better performance in comparison to empirical potentials.

As far as our literature search is concerned, ML potential-based investigations on the Au-Li binary system have not yet been reported.
Given their ability to reliably reproduce the different $ab~initio$ quantities at lower computational cost, ML potentials may be able to offer some resolutions on the above-mentioned discrepancies in this system.
Despite its successes in predicting alloy properties, ML potentials are not without challenges.
One such challenge is the proper generation of structural datasets needed for training the NNP, which is often achieved through $ab~initio$ molecular dynamics (AIMD) simulations.
The reliability of NNP strongly depends on the randomness of structural features of the dataset, which in turn requires long simulation runs.
To the best of our knowledge, a robust method of generating and preparing datasets for NNP training has not yet been developed.

In the present study, we constructed an NNP for the Au-Li binary system in an efficient and inexpensive way of structural dataset generation, which we used to investigate its alloying properties.
The accuracy of the NNP was corroborated by calculating the equation of state and phonon dispersions of some representative Au-Li alloy structures and comparing them with DFT results.
From the mixing energies of Au$_{1-x}$Li$_x$, we found that there were several structures at various alloy compositions that lay on or slightly above the convex hull, which could explain the discrepancies among the previously reported alloy phases.
We also found a new stable Au-Li alloy phase that has not yet been reported in the literature.
In conjunction, we present a new and efficient method of structural dataset generation for NNP training, which allows an inexpensive construction of accurate NNPs.

This paper is organized as follows.
Section~\ref{Method} describes the computational conditions of DFT and NNP, and the procedure to construct the NNP.
In Sec.~\ref{Results}, we present the results of predicted physical properties using the constructed NNP.
The stable phase search and alloying process of the phase separated structure are also provided in Sec.~\ref{Results}.
Finally, conclusions are presented in Sec.~\ref{Conclusions}.

\section{Methodology}
\label{Method}

\subsection{DFT calculations}
\label{Method: DFT}

\begin{figure}
\includegraphics[bb=0 0 3413 3083, width=0.5 \textwidth]{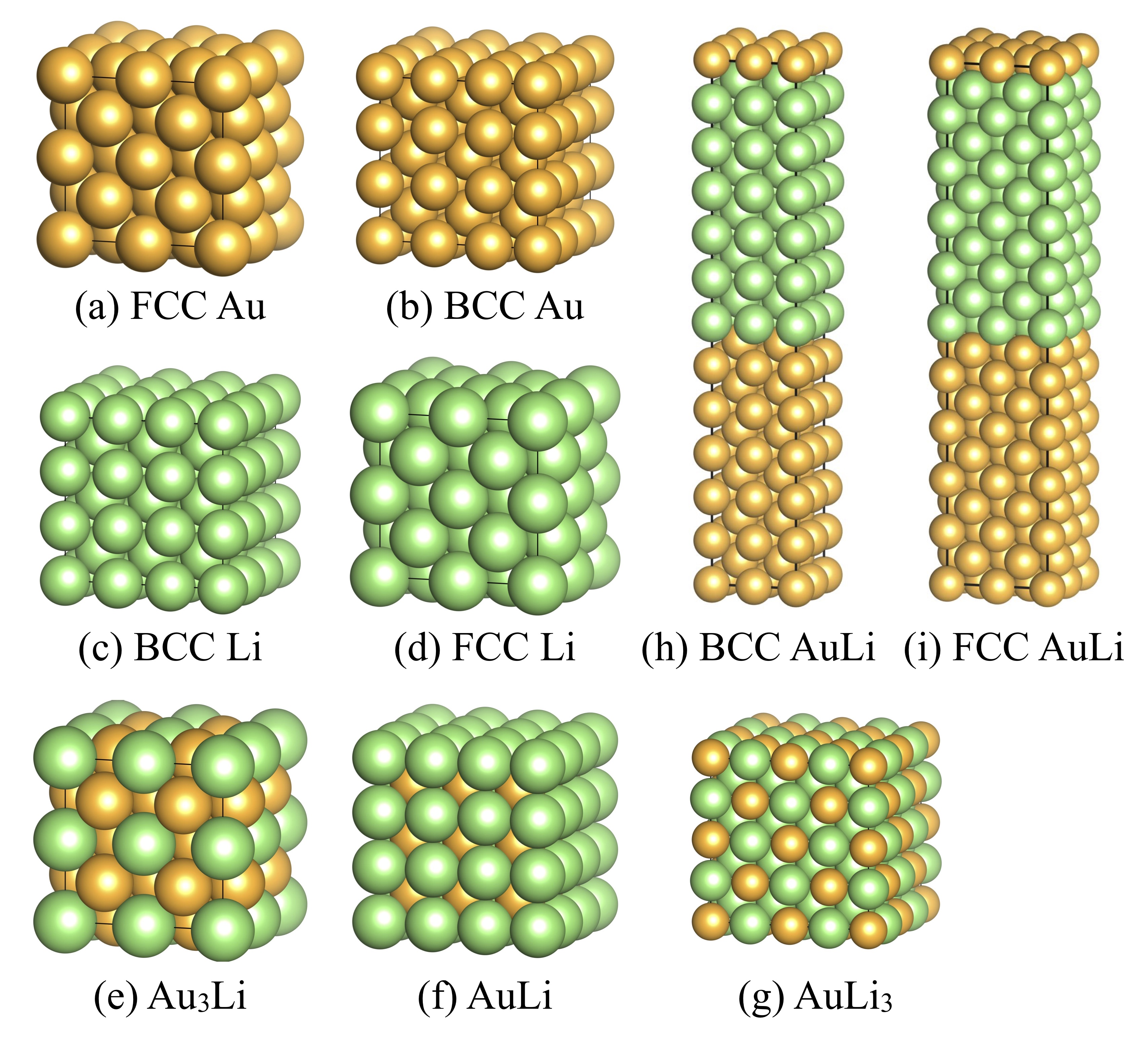}
\caption{
Base structures of (a, b) FCC and BCC Au, (c, d) BCC and FCC Li, (e) FCC Au$_3$Li, (f) BCC AuLi, (g) rocksalt AuLi$_3$, and (h, i) BCC and FCC Au$_{0.5}$Li$_{0.5}$ superlattices, used in the training dataset.
Structures are visualized using the VESTA package \cite{Momma-jac-2011}.
}
\label{structures}
\end{figure}

We firstly performed DFT-based AIMD calculations using the following structures to generate a temporal structural dataset for NNP construction: face centered cubic (FCC) Au, body centered cubic (BCC) Au, BCC Li, FCC Li, ordered FCC Au$_3$Li (32 atoms/supercell), BCC AuLi (54 atoms/supercell), and rocksalt AuLi$_3$ (128 atoms/supercell), as well as Au/Li superlattices of BCC Au$_{0.5}$Li$_{0.5}$ (96 atoms/supercell) and FCC Au$_{0.5}$Li$_{0.5}$ (160 atoms/supercell).
The supercells of these structures are shown in Fig.~\ref{structures} schematically.

We carried out constant temperature ($NVT$-ensemble) AIMD simulations with 1 fs time step for 1 ps.
The initial temperature was set to 300 K and linearly increased up to 2000 K (at 1.7 K/fs) to obtain largely random structural features.
Considering the high computational costs of AIMD calculations, further structural generations was conducted using NNP-based MD simulations and subsequent static DFT calculations were performed for the extracted structures.

For DFT calculations, we used the generalized gradient approximation with the Perdew-Burke-Ernzerhof functional \cite{PBE}, the plane wave basis set (500 eV cutoff energy), and the projector augmented wave method \cite{Bloechl-PAW}. 
Brillouin zone integration was performed using the sampling technique of Monkhorst and Pack ($5 \times 5 \times 5$ and $5 \times 5 \times 1$ sampling meshes for cubic cells and rectangular cells, respectively) \cite{Monkhorst-Pack}.
We used Vienna Ab initio Simulation Package (VASP) software \cite{VASP1, VASP2} for all the DFT calculations.

\subsection{NNP construction}

We adopted high-dimensional NNP \cite{Behler-PRL-2007} to describe the interatomic potential of Au and Li atoms.
In this scheme, the information of atomic structures is encoded as the values of symmetry functions (SFs) and is related to the energy contribution of each atom, $E_i$, via the neural network.
The total energy of the system is then described by the sum over the energy contributions of the respective atoms ($E = \sum_{i} E_{i}$).
Subsequently, the forces along the atomic coordinates $\alpha = x, y, z$ can be expressed as $F_{\alpha_{i}} = - \partial E / \partial \alpha_{i}$ = $-\sum_{\nu} \partial E / \partial G_{\nu} \times \partial G_{\nu} / \partial \alpha_{i}$, where $\nu$ labels the type of SFs.
We used the radial ($G_2$) and angular ($G_3$) SFs in the following forms.
\begin{equation}
G_{2}^{i} = \sum_{j \ne i} {\rm e}^{-\eta (R_{ij} - R_{s})^2 } f_{\rm c}(R_{ij}),
\label{sf2}
\end{equation}
\begin{equation}
\begin{split}
G_{3}^{i} = 2^{1-\zeta} \sum_{j \ne i} \sum_{k \ne i, j} (1 + & \lambda {\rm cos} \theta_{ijk} )  {\rm e}^{-\eta (R_{ij}^2 + R_{ik}^2) } \\
& f_{\rm c}(R_{ij}) f_{\rm c}(R_{ik}),
\end{split}
\label{sf3}
\end{equation}
where $\eta$ and $\zeta$ are the width parameters;
$\lambda$ and $R_{s}$ determine the peak positions;
$R_{ij}$ and $R_{ik}$ are the atomic distances of atom $i$ with $j$ and $k$, respectively; and
$\theta_{ijk}$ is the angle consisted of atoms $i$, $j$, and $k$ at the vertex $i$.
$f_{\rm c}$ is the cutoff function given by
\begin{equation}
f_{\rm c} (R_{ij}) = \begin{cases}
\cfrac{ [ {\rm cos} ( \frac{ \pi R_{ij} }{ R_{\rm c} } ) + 1 ] }{2} & (R_{ij} \leq R_{\rm c})  \\
0 & (R_{ij} > R_{\rm c})
\end{cases},
\label{cutoff}
\end{equation}
where $R_{\rm c}$ is the cutoff distance.

The set of hyper-parameters used in the present study are shown in Section S1 and Table S1 in Supplementary Materials.
We used 6 and 18 kinds of radial and angular SFs, respectively, for each elemental combination.
The combinations for Au atoms include Au-\{Au, Li\} and Au-\{Au-Au, Au-Li, Li-Li\} for radial and angular SFs, respectively.
Similarly, the combinations of SFs for Li atoms include Li-\{Au, Li\} and Li-\{Au-Au, Au-Li, Li-Li\}.
Thus, the total number of input nodes (i.e., values of SFs) is $2 \times 6 + 3 \times 18 = 66$ per atom. 
The SF values were normalized within the range of [$-$1:1] prior to passing through the NN.

Based on the total energies and atomic forces obtained by the DFT calculations, we optimized the weight parameters using the limited-memory Broyden-Fletcher-Goldfarb-Shanno (l-BFGS) algorithm \cite{Morales-2011} with the following loss function.
\begin{equation}
\begin{split}
\Gamma ({\bf w}) = & 
\alpha \sum^{N_{\rm train}}_{i=1} (E^{\rm NNP}_i - E^{\rm DFT} _i )^2 \\
& + \beta \sum^{N_{\rm train}}_{i=1} \{  \sum^{3n_i}_{j=1} (F^{i, {\rm NNP}}_j - F^{i, {\rm DFT}}_j )^2  \},
\end{split}
\label{loss}
\end{equation}
where ${\bf w}$ is the weight parameter vector.
$\alpha$ and $\beta$ determine the contribution ratio of energies and atomic forces into the loss function.
In the beginning of the NNP training, we evaluated 50 different, randomly assigned weight parameter sets, in which each weight value was chosen within the range of [$-$1:1].
Afterward, we started l-BFGS training with the smallest error weight set using energy differences, i.e., $\alpha = 1.0$ and $\beta = 0$.
Subsequently, we continued the training using both energy and atomic force differences using the setting $\alpha = 0.5$ and $\beta = 0.5$.

\subsection{Structural dataset generation}

\begin{figure}
\includegraphics[bb=0 0 3189 1899, width=0.5 \textwidth]{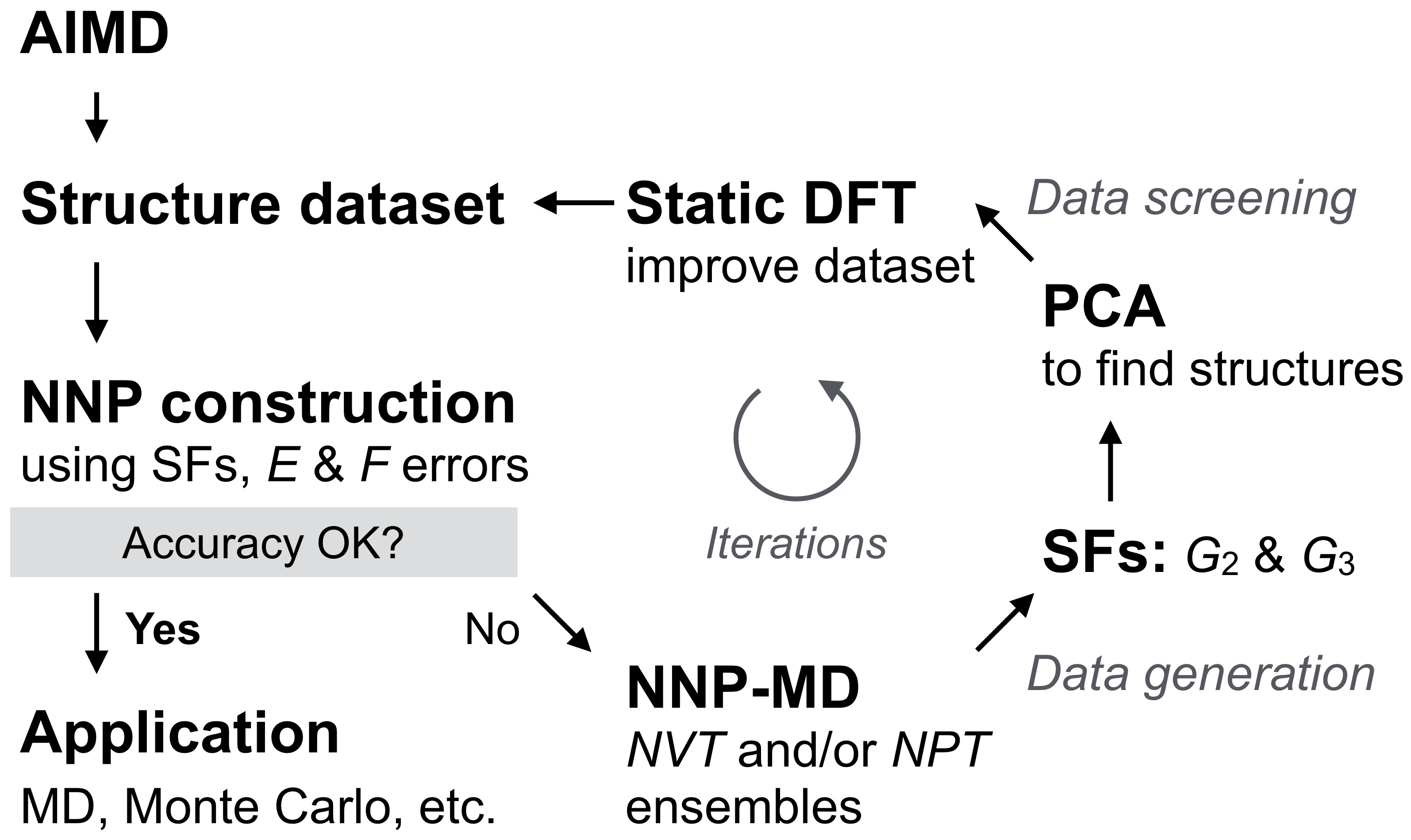}
\caption{
A schematic description of the workflow of NNP construction and dataset generation.
}
\label{flow}
\end{figure}

\begin{figure}
\includegraphics[bb=0 0 3036 3853, width=0.5 \textwidth]{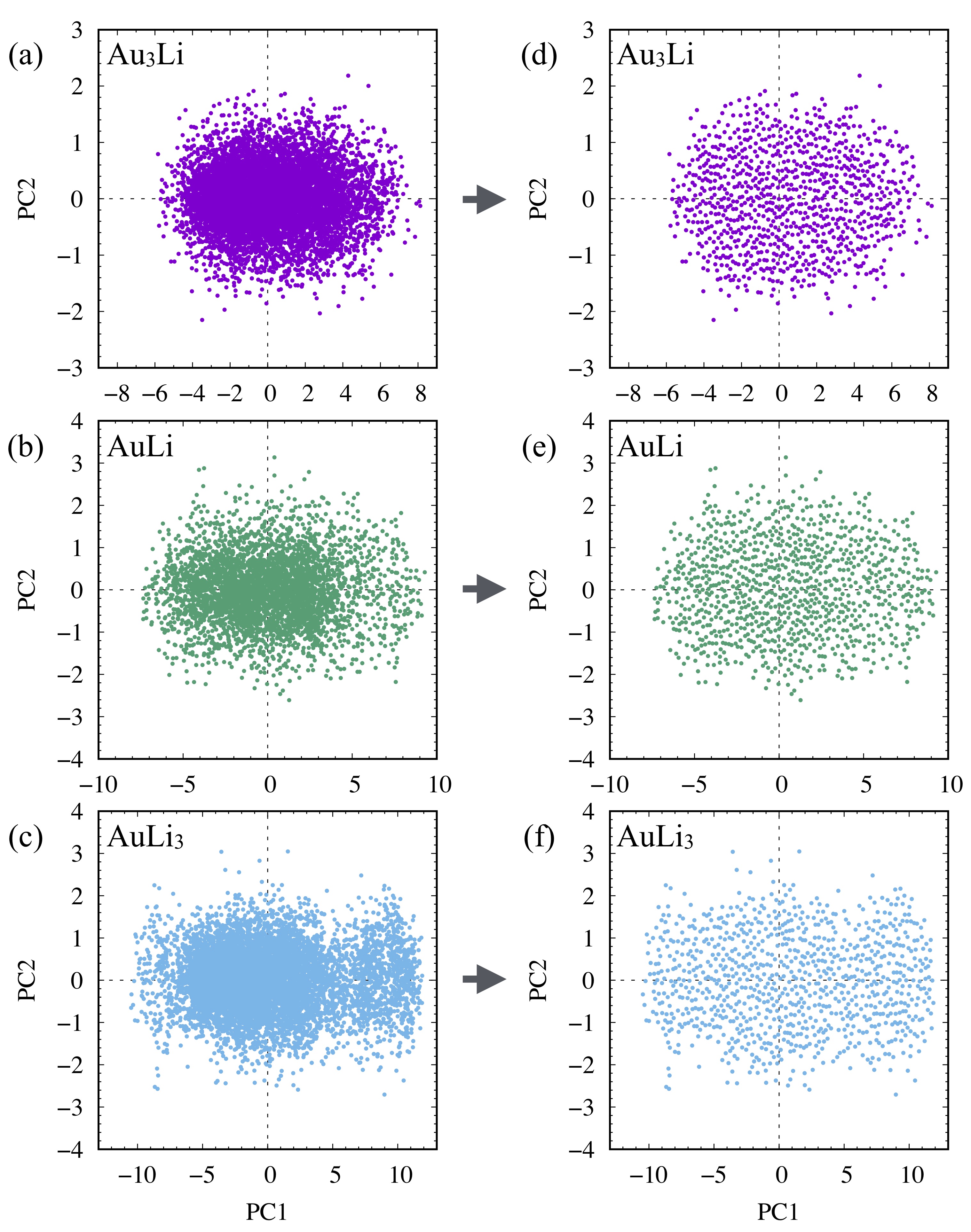}
\caption{
Principal component plots of (left) a large number of structures generated by NNP-MD simulations and (right) a reduced number of structures.
(a, d) Au$_3$Li, (b, e) AuLi, and (c, f) AuLi$_3$ are shown as examples.
}
\label{pca}
\end{figure}

\begin{figure}[h]
\includegraphics[bb=0 0 2396 1858, width=0.5 \textwidth]{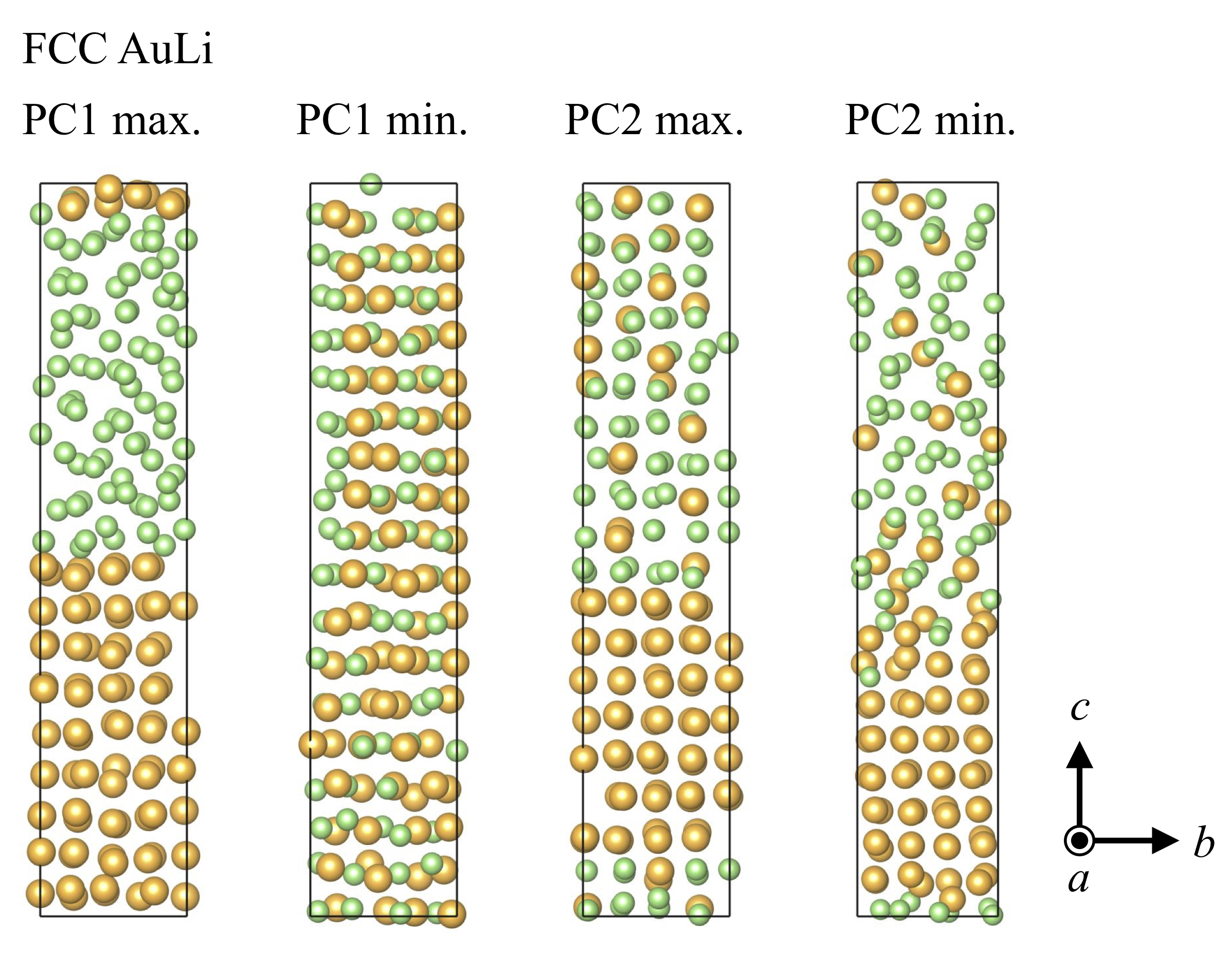}
\caption{
Structures of rectangular FCC Au$_{0.5}$Li$_{0.5}$ at the boundaries of principal component (PC) values (cf., Fig.~\ref{pca}).
}
\label{pca_fcc}
\end{figure}

In this section, we will explain the efficient method of NNP construction and structural dataset generation that we developed for the present study.
Figure~\ref{flow} shows the workflow of our NNP construction.
First, we constructed a tentative NNP using the structures obtained by the AIMD calculations (please refer to Section \ref{Method: DFT} for the computational conditions).
Although this tentative NNP is usually less accurate and insufficient for practical use, it was enough to be used for generating additional structures that were added to the training dataset.
To obtain these additional structures, we coupled our NNP with Large-scale Atomic/Molecular Massively Parallel Simulator (LAMMPS) software \cite{LAMMPS1, LAMMPS2} using our homemade interface and carried out $NVT$- and $NPT$-ensembles MD calculations for the structures shown in Fig.~\ref{structures}.
In the present study, we performed 10-100 ps NNP-MD with a 1 fs time step.
The temperatures were set from 300-2100 K at 300 K intervals for the Au case, 300-600 K at 100 K intervals for the Li case, and 300-1200 K at 100 K intervals for the Au-Li alloy cases.
The maximum temperatures were chosen based on being several hundreds of Kelvin higher than the materials' melting temperatures.
It is worth noting that rough NNP-MD simulations are often terminated by numerical errors due to structure collapses as mentioned in Ref.~\cite{Jeong-JPCC-2018, Lee-CPC-2019}, wherein the training technique that can construct accurate and numerically stable NNPs was developed.

To select only those structures that have different structural features from the ones already included in the collected dataset, we performed principal component analysis (PCA) based on the SFs of each structure, where $66 \times N$-dimensional (where $N$ = number of atoms/structure) local atomic/structural features were reduced to 2-dimensional values, i.e., first and second principal components (PC1 and PC2).
Having similar PC values means that those structures have similar structural features.
In fact, structures resembling each other can be found frequently in MD trajectories.
We calculated SFs for 1\% of structures extracted from the NNP-MD trajectories.
Then, we chose a specific number of structures by calculating the distances of neighboring points so that the selected PC values were well-scattered (please refer to Section S2 in Supplementary Materials for more details).
Figure~\ref{pca}(a-c) shows the PC plots of many of the structures of Au$_3$Li, AuLi, and AuLi$_3$, respectively.
The plots of the extracted structures are shown in Fig.~\ref{pca}(d-f).
The well-scattered plots in all the cases presented suggest that the dataset contained various structures and atomic features (please also refer to Fig. S2 in Supplementary Materials for resulting PC plots).
As an example, Figure~\ref{pca_fcc} shows the structures corresponding to the maximum and minimum PC values in rectangular FCC Au$_{0.5}$Li$_{0.5}$.
The maximum PC1 corresponds to the phase separated structure.
Conversely, at the minimum PC1 value, the Au and Li atoms were completely mixed.
Both the maximum and minimum PC2 show the partial mixture of Au and Li atoms, which is reasonable because their PC1 values were located near the origin.
The structures wee more disordered as the PC2 value decreased, although the difference is rather ambiguous compared to PC1 cases.
After PCA, we subsequently performed static DFT calculations for the new structures to obtain the total energies and atomic forces.
These were then used to update our dataset and retrain our NNP.
We performed this cycle iteratively until we could achieve an accurate NNP.

We generated 8696 structures in total, including 1973 Au, 1900 Li, 949 Au$_3$Li, 992 AuLi, 929 AuLi$_3$, 971 rectangular BCC Au$_{0.5}$Li$_{0.5}$, and 982 rectangular FCC Au$_{0.5}$Li$_{0.5}$ structures.
However, we found that at low-Au-concentration alloys, the calculated mixing energies using the NNP seriously deviated from the DFT verifications.
Therefore, we further added 589 Au dilute alloy structures, i.e., Au$_1$Li$_{31}$, Au$_1$Li$_{53}$, and Au$_1$Li$_{127}$, obtained by the procedure explained above.
Finally, our dataset containing 9285 structures covered the entire Au-Li space, suggesting the versatility of the constructed NNP.

We note that Artrith and Behler proposed a method to find structures worth adding to the dataset in the light of structural diversity using NNPs \cite{Artrith-PRB-2012}.
In their method, NNPs with different network architectures constructed by the same dataset are used to predict energies for a large number of structures (generated for instance by NNP-MD).
A large difference in predicted energies of a same structure among NNPs indicates that its atomic features are missing in that dataset.
In addition, Li and Ando applied this scheme to the $on$-$the$-$fly$ sampling technique \cite{Li-JCP-2019}.
Compared with these previous methods, the present approach needed to construct only a single NNP.
The lower computational costs for the training process would be the advantage against the above methods.

\section{Results \& Discussion}
\label{Results}

\subsection{Au-Li binary system NNP}

Using the 9285 structures mentioned in the previous section, we constructed NNPs for the Au-Li binary system.
We examined cutoff distances of 5, 7, and 9 \AA~with the NNs consisting of 2 hidden layers and 10 or 20 nodes per hidden layer.
The obtained root-mean-square-error (RMSE) values are summarized in Section S1 and Table S2 in Supplementary Materials.
We used a randomly chosen 10\% of the structures in the dataset as the test/validation data.
We found that all the constructed NNPs predicted nicely both the DFT total energies and atomic forces.
The RMSE values are comparable to or even better than the previous studies \cite{Behler-PRL-2007, Li-JPC-2017, Onat-PRB-2018, Jeong-JPCC-2018, Minamitani-APE-2019, Artrith-PRB-2011}.
Note that the good prediction performance of the NNP with a cutoff distance of 5 \AA~suggests the insignificance of long-range interaction in Au-Li systems.
Considering the tradeoff between reliability of longer cutoff distances and computational costs, we used the 7 \AA~cutoff distance and [66-10-10-1] network structure hereafter.
The RMSE of total energies and atomic forces in this setting were 1.53 meV/atom and 23.1 meV/\AA~for the training dataset and 1.46 meV/atom and 22.9 meV/\AA~for the test dataset, respectively.

\begin{figure}[h]
\includegraphics[bb=0 0 2849 2773, width=0.5 \textwidth]{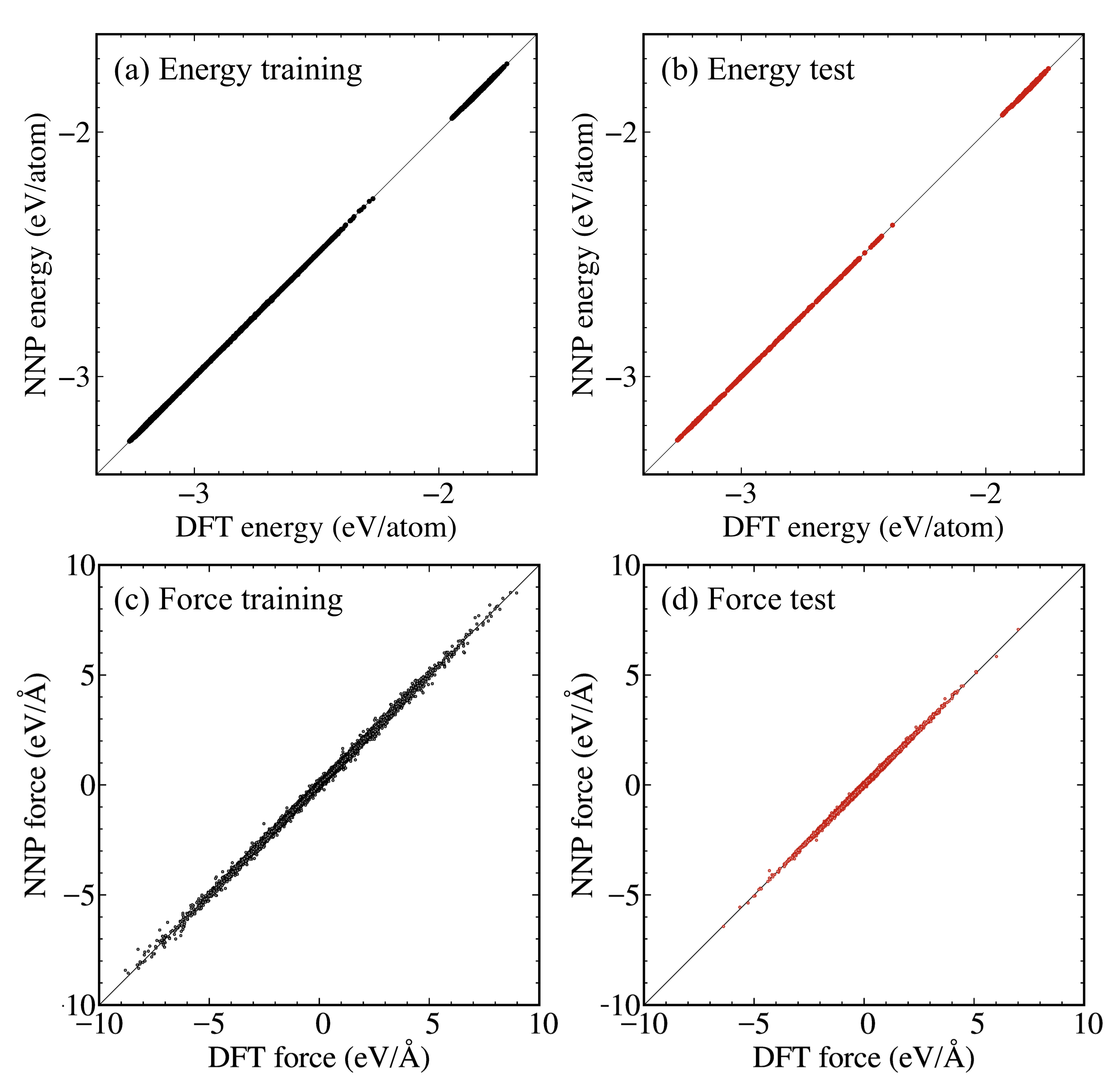}
\caption{
Comparison between DFT and NNP on the total energies of (a) training and (b) test sets.
(c) and (d) show the comparison of the atomic forces for training and test sets, respectively.
}
\label{predict}
\end{figure}

Figure~\ref{predict} shows the comparison between DFT and NNP on the total energies and atomic forces.
Both total energies and atomic forces are aligned along the diagonal lines, suggesting that the constructed NNP accurately predicted all structure types considered, i.e., Au, Li, Au$_3$Li, AuLi, AuLi$_3$, and dilute-Au alloy.
In addition, the learning curves of the total energies and atomic forces are shown in Fig.~S1.
The mean-square-error (MSE) values monotonically decreased for both total energies and atomic forces, which indicates that no overfitting occured.
Note that the preceding 968 ($\alpha = 1.0, \beta = 0.0$) and 7933 ($\alpha = 0.5, \beta = 0.5$) iterations were performed ahead of the last training curves shown in Fig.~S1.

\subsection{Lattice constants and phonon properties}

\begin{figure}[h]
\includegraphics[bb=0 0 2841 1362, width=0.5 \textwidth]{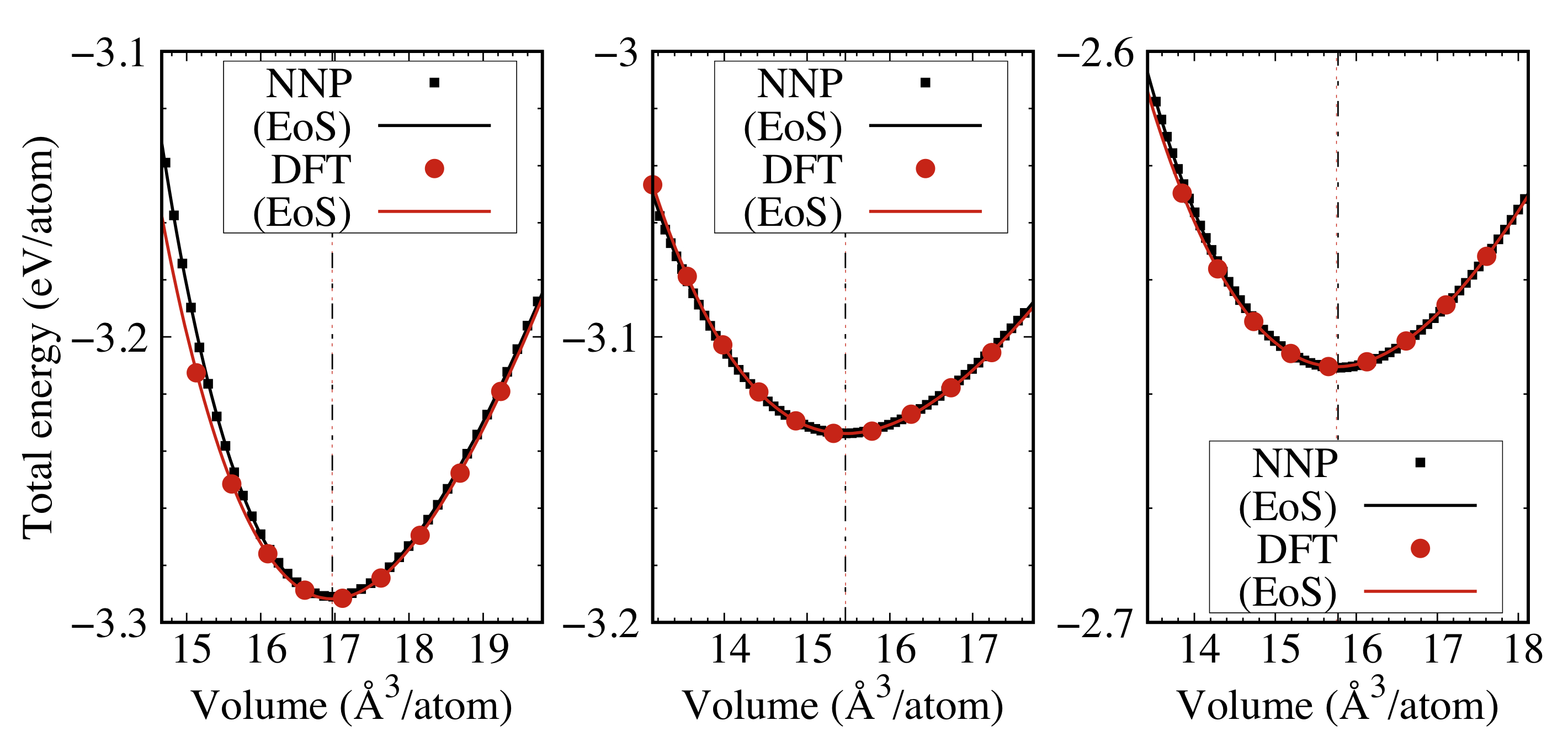}
\caption{
Comparison of total energies as a function of volume per atom/lattice constant obtained by DFT and NNP for (left) bulk FCC Au$_3$Li, (center) BCC AuLi, and (right) rock-salt AuLi$_3$.
The solid lines indicate the fitted curves of equation of state.
The vertical dotted line and dashed-dotted line correspond to the minimum volume per atom/lattice constant of DFT and NNP, respectively.
Note that the two lines are almost overlapped.
}
\label{lattice2}
\end{figure}

To verify the accuracy of the constructed NNP, we performed the lattice constant optimization using Au and Li, as well as Au$_3$Li, AuLi, and AuLi$_3$ ordered alloys.
Figures~\ref{lattice2} and S3 show the profiles of total energy as a function of volume per atom obtained by DFT and NNP calculations, together with the fitted Birch-Murnaghan equation of state (EoS).
The vertical dashed-dotted and dotted lines correspond to the volume at the energy minima obtained by DFT and NNP, respectively.
The DFT and NNP results agreed well on the minimum volume and energy: the maximum differences were smaller than 0.1 \AA$^3$/atom and 3.78 meV/atom.
The bulk moduli were also accurately predicted within a 10\% difference, wherein the deviations mainly originated at the compressed structures.
Detailed results on fitting to the Birch-Murnaghan equation are given in Section S3 and Table S3 in Supplementary Materials.

\begin{figure}[h]
\includegraphics[bb=0 0 3739 1457, width=0.5 \textwidth]{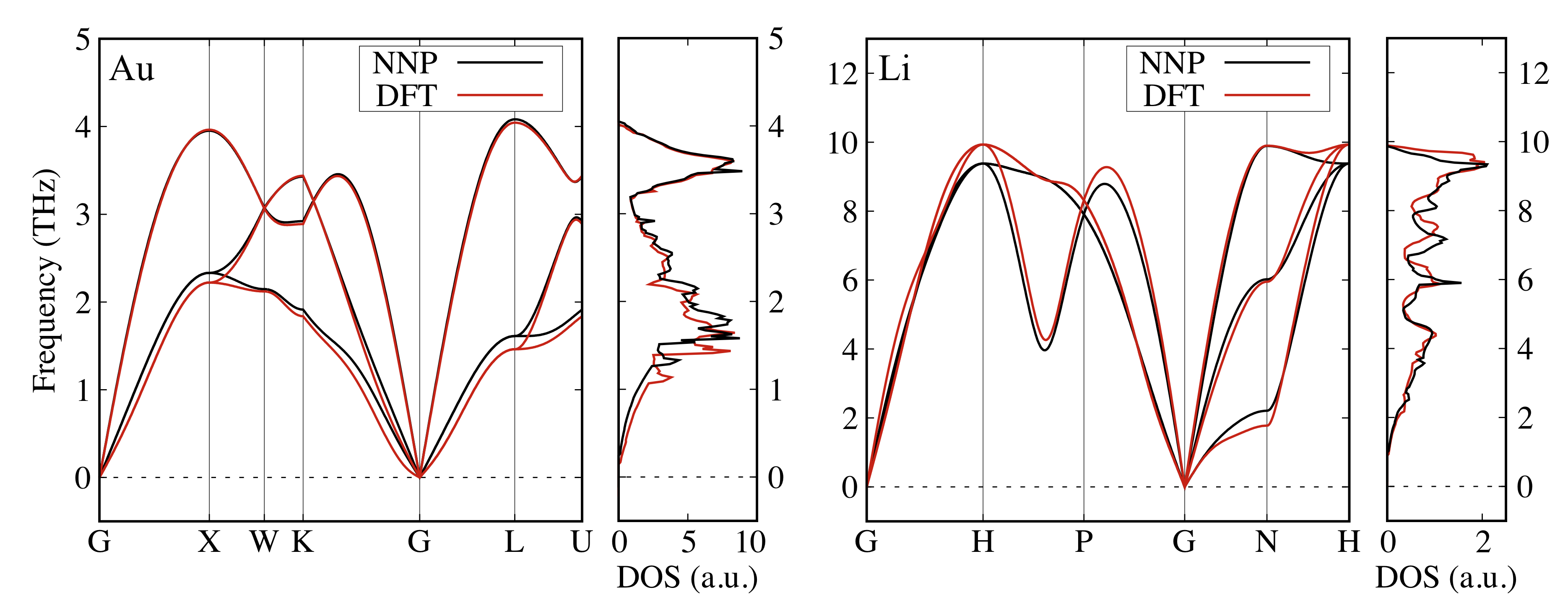}
\caption{
Comparison of phonon bands and densities of states obtained by DFT and NNP for (left) FCC Au and (right) BCC Li.
}
\label{band1}
\end{figure}

\begin{figure*}[t]
\includegraphics[bb=0 0 5493 1450, width=1 \textwidth]{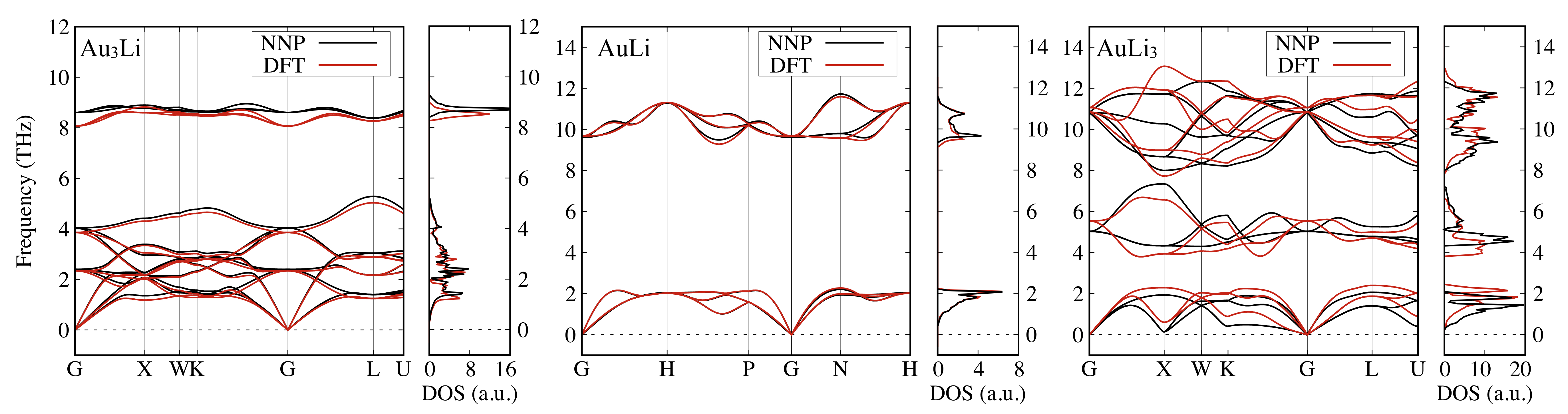}
\caption{
Comparison of phonon bands and densities of states obtained by DFT and NNP for (left) FCC Au$_3$Li, (center) BCC AuLi, and (right) rocksalt AuLi$_3$ ordered alloys.
}
\label{band2}
\end{figure*}

Next, we performed phonon calculations for FCC Au, BCC Li, FCC Au$_3$Li, BCC AuLi, and rocksalt AuLi$_3$ structures based on DFT and NNP using phonopy software \cite{Togo-SM-2015}.
We used the same supercell size as shown in Fig.~\ref{structures} and 0.05 \AA~displacements to calculate the force constants.
Note that we used 0.02 \AA~displacements for the AuLi$_3$ case to circumvent the appearance of imaginary frequencies around the X-point.

Figure~\ref{band1} shows the phonon band structures and the corresponding phonon densities of states (DOS) of FCC Au and BCC Li calculated by DFT and NNP.
We found excellent agreement between the DFT- and NNP-obtained phonon modes and the peak structures of phonon DOS.
These phonon features could also be found in other studies \cite{Lynn-PRB-1973, Corso-JPCM-2013, Xu-PRL-2014}.
Similarly, Figure~\ref{band2} shows the phonon bands and the corresponding phonon DOS of FCC Au$_3$Li, BCC AuLi, and rocksalt AuLi$_3$.
The acoustic modes have been well predicted by NNP while the optical modes of Au$_3$Li show a slight discrepancy around the $\Gamma$-point.
However, all the phonon modes match in the AuLi case.
This may be attributed to the large number of Au:Li = 1:1 structures included in the training dataset.
The phonon bands of AuLi$_3$ show rather large discrepancies, which is probably due to the relatively sparse sampling of this composition (cf., Fig.~S2 especially around the origin).
Nonetheless, the phonon dispersion and energy levels were still well reproduced.

It is worth emphasizing that the phonon band structures suggest the high stability of these compounds.
Note that we obtained imaginary frequencies for BCC Au and FCC Li phonon bands, which would be reasonable considering the instability of these structures (not shown).
This verifies the accuracy of our NNP as well as its capability to reproduce lattice parameters and phonon properties.

\subsection{Phase stability of Au-Li compounds}

\begin{figure}
\includegraphics[bb=0 0 5343 2587, width=0.5 \textwidth]{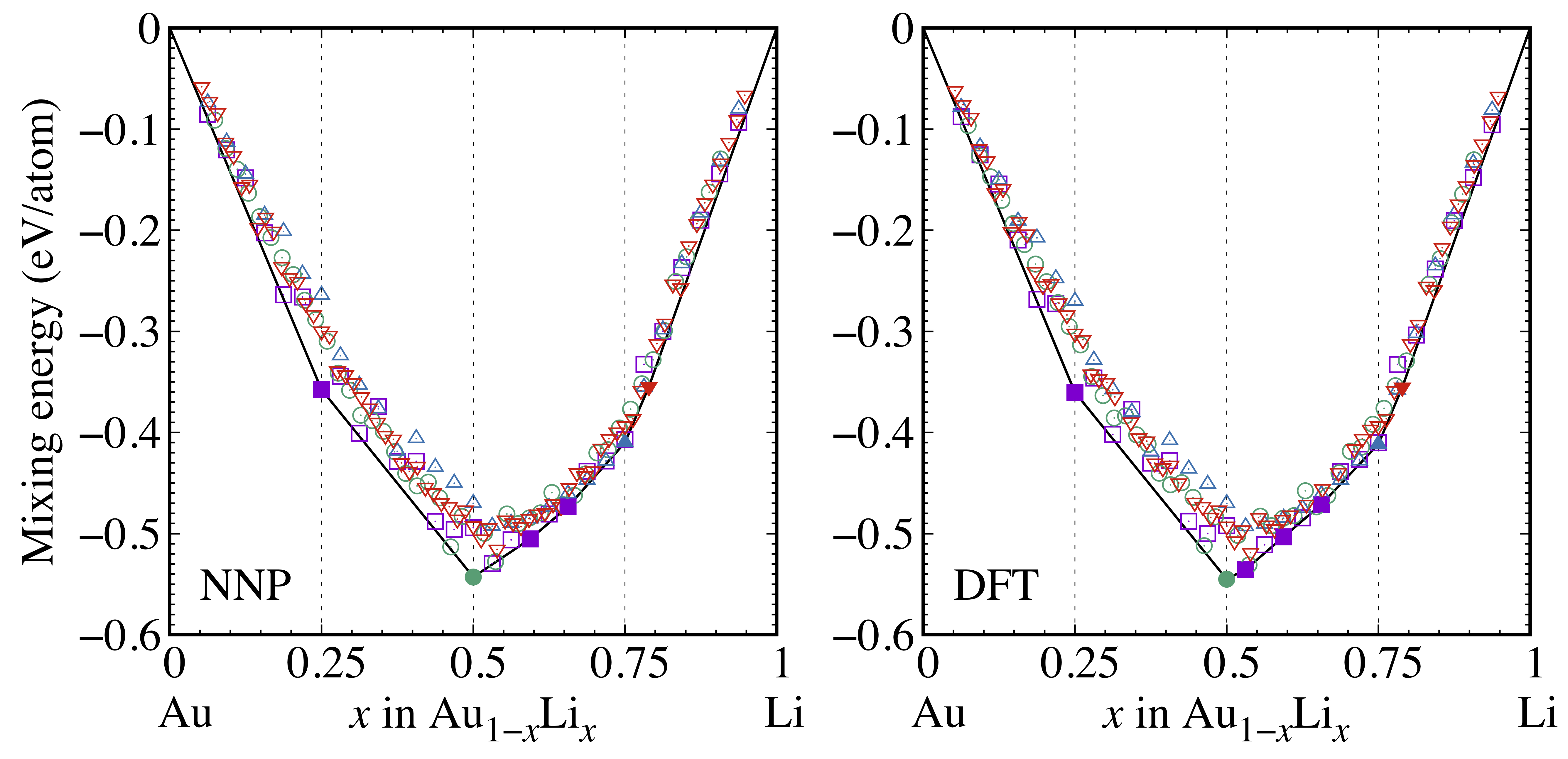}
\caption{
Calculated mixing energy of Au$_{1-x}$Li$_x$ using (left) NNP and (right) DFT.
Filled symbols show the convex hull points, which are connected by the solid lines.
}
\label{mixing}
\end{figure}

\begin{figure}[h]
\includegraphics[bb=0 0 5165 1551, width=0.5 \textwidth]{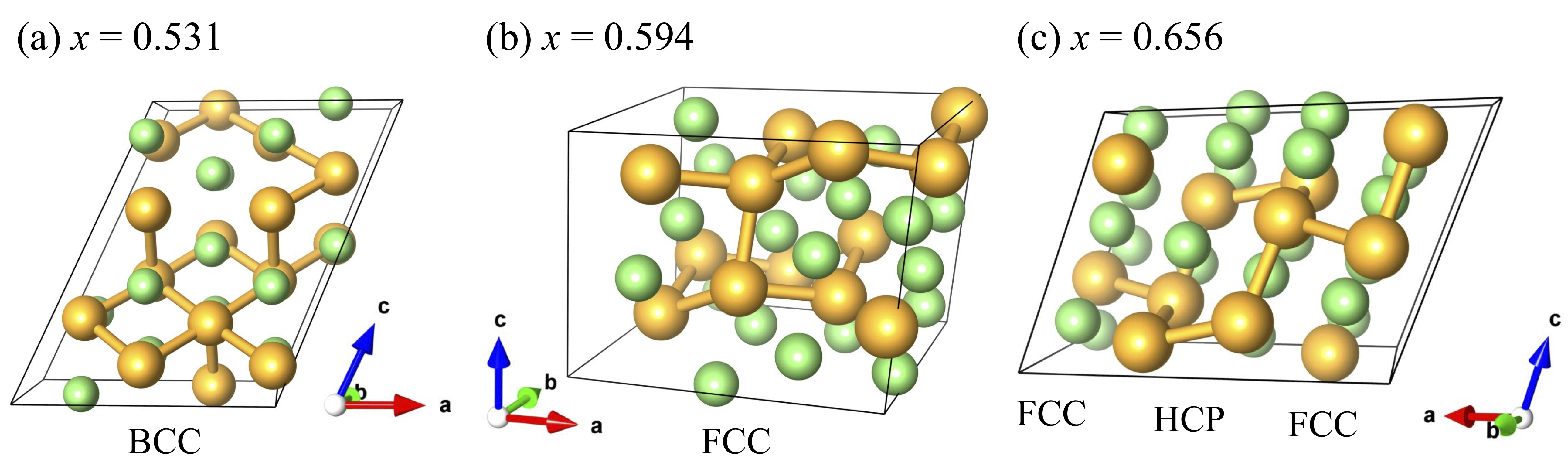}
\caption{
Structural images of Au$_{0.469}$Li$_{0.531}$, Au$_{0.406}$Li$_{0.594}$, and Au$_{0.344}$Li$_{0.656}$, which consist of the convex hull points.
}
\label{new}
\end{figure}

\begin{figure*}[t]
\includegraphics[bb=0 0 4909 1325, width=1 \textwidth]{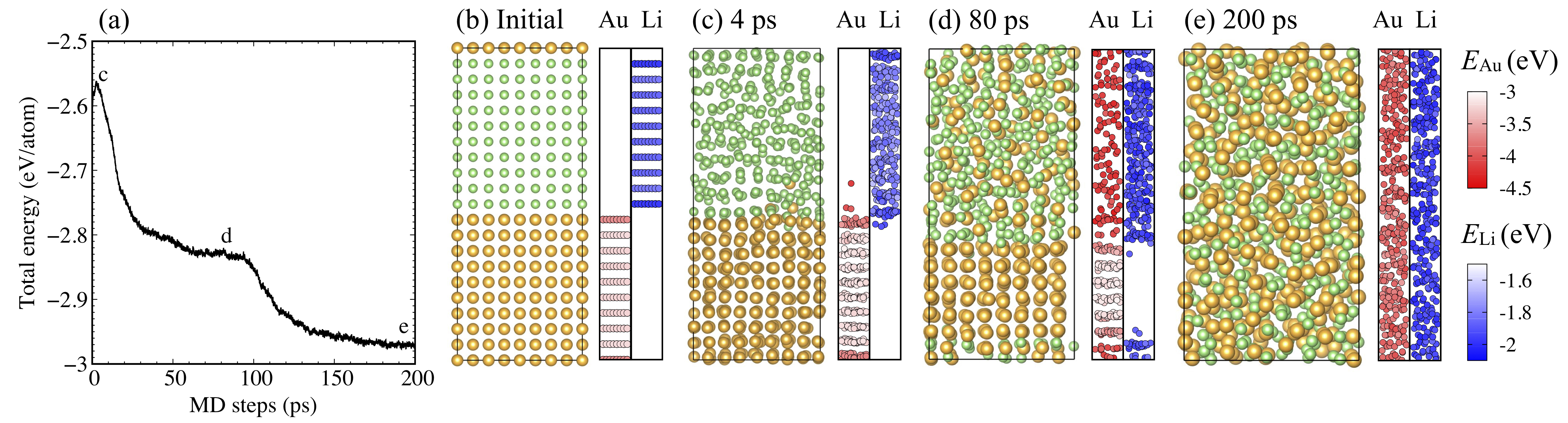}
\caption{
(a) Total energy profile of the NNP-MD simulation, and its snapshots of (b) initial, (c) 4 ps, (d) 80 ps, and (e) 200 ps.
The atomic energies of Au and Li are separately shown in the right-hand-side of each snapshot using relative atomic positions.
The colors correspond to the magnitude of the atomic energies.
}
\label{long_fcc}
\end{figure*}

As mentioned in the Introduction, the stable Au-Li alloy phases differ among the previous research \cite{Pelton-BAPD-1986, Bach-EA-2015, Bach-CM-2016, Yang-JACS-2016}, which arises from the intrinsic complexity of miscible alloy materials and the expensive computational costs.
Here, utilizing the constructed NNP, we examined the mixing energies of Au-Li with fine composition grids.
We used the ordered structures of Au$_3$Li, AuLi, AuLi$_3$, and Au$_4$Li$_{15}$ as the base structures (cf., Fig.~\ref{structures} and Fig.~S4 in Supplementary Materials), and Au and Li atoms were randomly replaced for the sake of compositional variations.
Note that we replaced Au and Li at intervals of 1 and 4 atoms, respectively, for Au$_3$Li, AuLi, Au$_4$Li$_{15}$, and AuLi$_3$ base structures.
Each structure was first equilibrated by an $NPT$ ensemble MD simulation for 200 ps at $T =$ 500 K.
Our choice of $T =$ 500 K is justified by the finding that this value seemed high enough to facilitate the migration of atoms to reach stable configurations.
This was then followed by an $NVT$ and subsequently $NVE$ ensembles MD simulations for 200 ps each with the same temperature.
Lastly, we fully optimized the structures (including lattice vectors and volume optimization) using the conjugate gradient method with 10$^{-3}$ meV and 1 meV/\AA~of the total energy and atomic forces convergence criteria, respectively.
As verification, we checked the total energies of resulting structures using static DFT calculations.
The mixing energies were calculated by $E_{\rm mix} = E({\rm Au}_{1-x}{\rm Li}_{x}) - \{ (1-x)E({\rm Au}) + xE({\rm Li}) \}$, where $E$ is the energy per atom of the composition designated in the parentheses.

Figure~\ref{mixing} shows the calculated mixing energies using the NNP and static DFT calculations.
The squares, circles, triangles, and inverse triangles correspond to Au$_3$Li, AuLi, AuLi$_3$, and Au$_4$Li$_{15}$ base structures, respectively.
The filled symbols indicate the so-called convex hull points, which are connected by the solid lines.
The resulting mixing energies show good agreement between NNP and DFT over the whole compositional space.
While one exception exists at the convex hull point with $x = 0.531$, the extremely small energy difference of the order of 5.80 meV/atom at this point is indistinguishable.
From the mixing energy profiles, we found that Au$_3$Li, AuLi, AuLi$_3$, and Au$_4$Li$_{15}$ consisted of the convex hull, as is the case in the MP.
In addition, we found Au$_{0.469}$Li$_{0.531}$, Au$_{0.406}$Li$_{0.594}$, and Au$_{0.344}$Li$_{0.656}$ as the hull points.
The structures of these three phases are shown in Fig.~\ref{new}.
The common neighbor analysis method \cite{Stukowski-MSMSE-2012} identified that Au$_{0.469}$Li$_{0.531}$ and Au$_{0.406}$Li$_{0.594}$ have BCC and FCC crystal structures, respectively.
Conversely, Au$_{0.344}$Li$_{0.656}$ had the laminated structure of two layers of FCC and HCP along the $a$-axis.
The radial distribution functions (RDFs) clearly show that Au$_{0.469}$Li$_{0.531}$ simply replaced a few Au and Li atoms, keeping the AuLi crystal structure.
Although the first peaks of Au$_{0.406}$Li$_{0.594}$ and Au$_{0.344}$Li$_{0.656}$ were located close to that of Au$_3$Li, the attenuations of subsequent peaks are indicative of their rather amorphous-like structures.
Note that the RDFs are shown in Fig.~S5 in Supplementary Materials.
In addition to the 7 convex hull points, we also found several points located slightly above the hull over the entire compositional space.
These compositions may appear in a specific experimental condition as meta-stable phases.

We also performed the mixing energy calculations for the previously reported stable phases, i.e., Au$_2$Li, Au$_5$Li$_3$, Au$_2$Li$_3$, Au$_3$Li$_5$, and AuLi$_2$, starting from the provided space group information \cite{Bach-CM-2016, Yang-JACS-2016}.
The resulting mixing energies of Au$_5$Li$_3$, Au$_2$Li$_3$, and AuLi$_2$ were located at 3.38, 6.04, and 4.97 meV/atom below the convex hull of Fig.~\ref{mixing}, respectively, while Au$_2$Li and Au$_3$Li$_5$ were 15.1 and 3.28 meV/atom higher than the convex hull, respectively, for the NNP case.
Contrastingly, for all five compositions, the mixing energies obtained by DFT were located below the convex hull (please refer to Section S4 and Fig. S6 in Supplementary Materials for details).
The mixing energies of Au$_2$Li, Au$_5$Li$_3$, Au$_2$Li$_3$, Au$_3$Li$_5$, and AuLi$_2$ were 1.58, 5.88, 4.16, 5.41, and 3.91 meV/atom lower than the convex hull lines at the corresponding composition ratio, respectively.
However, the subtle differences in mixing energies in the order of a few meV/atom could be easily buried by circumstances and/or numerical accuracy.
Thus, observing a part of the original phase is unavoidable.
The discrepancy among previous studies mentioned earlier can be understood from the above reasoning.
Note that while considering these phases, Au$_{0.469}$Li$_{0.531}$ still consisted of the convex hull point.

As demonstrated above, our constructed NNP can be used over the entire range of binary alloy compositions, despite our training dataset including only a limited number of Au-Li ratios, $x$.
The precise investigation of mixing energy reveals the compositions that were not present in the previous studies.
However, despite the fine intervals of $x$, some crystal structures, inaccessible by the systems used in this work, exist.
In fact, the five previously reported crystal structures were not covered in the search of Fig.~\ref{mixing}.
Further consideration, such as a different number of atoms per supercell, may enable us to draw a more accurate phase diagram.
We shall leave this to our future studies.

Next, to clarify the atomic-scale picture of the alloying process, we carried out NNP-MD ($NPT$ at $T =$ 500 K) simulations starting from the phase-separated structure, consisting of 10 atomic layers (32 atoms/layer) of FCC Au and Li using the Au lattice constant (Fig.~\ref{long_fcc}(b)).
Figure~\ref{long_fcc}(a) shows the total energy profile along the simulation time, and (b)-(e) show the MD snapshots together with the atomic energy profiles.
The colors indicate the magnitude of atomic energies along with the relative atomic positions.

We found that at 4 ps, a single Au atom dissolved into the bulk Li (Fig.~\ref{long_fcc}(c)).
This occured due to the slightly larger energy gain of Au dissolving into Li than the Li dissolving into Au, which is corroborated by the mixing energies (cf., Fig.~\ref{mixing}).
This was followed by continued Au alloying into the Li until around 80 ps, where the total energy reached a plateau (Fig.~\ref{long_fcc}(a)).
At this plateau, Li mixed with Au over the entire Li bulk region while 6 layers of Au remained intact (Fig.~\ref{long_fcc}(d)).
The cell volume gradually decreased by reducing the length of the $c$-axis, which is attributed to the smaller optimal volume of alloys (cf., Figs.~\ref{lattice2} and S3).
The alloying was reinitiated at ca. 100 ps with the emergence of a second plateau in the total energy profile.
At 200 ps, complete alloying and equilibration (cf., Fig.~\ref{long_fcc}(e)) was reached.

The solution of Au atoms into Li at the initial stage of allying is a probabilistic process, and it proceeds independently at the both Au-Li interfaces.
When multiple adjacent Au atoms dissolved into Li, the alloying of the entire interface started from the dissolved region.
This corresponds to the steep energy decrease from ca. 4 to 30 ps as shown in Fig.~\ref{long_fcc}(a). 
In the present case, the dissolution of the lower interface initially took place. 
The upper interface follows from the inflection point of the energy profile at ca. 12 ps. 
Note that the initial alloying process using 128 atoms/layer with 10 atomic layers of FCC Au and Li model is shown in Section S5 in Supplementary Materials.

In terms of atomic energies, atoms located at the interface had lower values (Au: $-3.71$ eV, Li: $-2.02$ eV) compared to those of bulk region (Au: $-3.28$ eV, Li: $-1.89$ eV) at the initial structure.
Atoms at the second layer from the interface, conversely, had higher values (Au: $-3.23$ eV, Li: $-1.84$ eV).
This shows that the Au-Li sharp interface will spontaneously form an alloy.
While thermal oscillation caused the atomic energy fluctuation at the bulk region, e.g., 41.0 meV at 4 ps and 51.3 meV at 80 ps, the interface atoms had much lower energies.
Once Au atoms mixed with Li atoms, the atomic energies decreased further.
A major part of atomic energy change by alloying was absorbed into the Au section.
We shall also leave this point for future work.

As demonstrated above, due to the high stability of Au-Li system, Au and Li atoms spontaneously mixed in the simulations within a few hundred picoseconds, which is a noticeably short period compared to the operating time of ionic devices.
This alloying property, in other aspects, leads to the stabilization of interfaces of secondary batteries using a Li anode (cf., e.g., Refs.~\cite{Kato-JPS-2016, Kato-SSI-2018}).
On the other hand, controlling the amount of Li introduced into Au must be a significant factor for developing devices using ion conduction, such as VolRAM \cite{Sugiyama-aplm-2017}.

\section{Conclusions}
\label{Conclusions}
We constructed a neural network potential (NNP) for Au-Li binary systems based on density functional theory (DFT) calculations.
We used a total of 9285 structures, including pure Au and Li, ordered Au$_3$Li, AuLi, and AuLi$_3$, dilute Au alloys, and Au/Li superlattices as our dataset.
We efficiently generated these structures using NNP-based MD simulations, where unique structures were identified by principal component analysis based on symmetry functions.
The predicted lattice parameters and phonon properties obtained by the NNP agree well with those from DFT calculations.
We also investigated the mixing energy of Au$_{1-x}$Li$_x$, showing consistency with the DFT results over a wide range of compositional space.
We successfully reproduced the convex hull points of Au$_3$Li, AuLi, AuLi$_3$ and Au$_4$Li$_{15}$, which are reposited in the Materials Project database as the stable phases.
Additionally, we found three new alloy compositions of Au$_{0.469}$Li$_{0.531}$, Au$_{0.406}$Li$_{0.594}$, and Au$_{0.344}$Li$_{0.656}$, consisting of the convex hull.
We further verified the previously reported phases of Au$_2$Li, Au$_5$Li$_3$, Au$_2$Li$_3$, Au$_3$Li$_5$, and AuLi$_2$ as stable phases.
For compositions where there are discrepancies in stability among previous studies, we found that a small energy change of several meV/atom can alter whether or not they become stable phases.
This is considered to be the origin of the discrepancies.
Considering a vast search space and marginal energy differences among alloy compositions, inexpensive and accurate simulations using machine learning interatomic potentials could be utilized for further studies. 
Finally, we examined the alloying process starting from the phase separated structure to the complete mixing phase using our relatively large system.
We found that at the initial stage of alloying, a single Au atom tends to dissolve into the Li region.
Furthermore, we determined that when multiple adjacent Au atoms dissolve into Li, the alloying of the entire Au/Li interface starts from the dissolved region.

In this study, we successfully constructed an accurate NNP via the proposed workflow.
This would help to accelerate applications of ML interatomic potentials toward more complicated systems.

Acknowledgements.
This work was supported by the
JST CREST Program ``Novel electronic devices based on nanospaces near interfaces" and
JSPS KAKENHI Grant Numbers 19H02544, 19K15397, 20K15013, 20H05285.
Some of the calculations presented here were performed using the computer facilities at
ISSP Supercomputer center and Information technology center, The University of Tokyo, and Institute for Materials Research, Tohoku University (MASAMUNE-IMR).
We would like to thank Editage (www.editage.com) for English language editing.


\bibliography{apssamp}

\end{document}